\documentclass{article}

\usepackage{microtype}
\usepackage{graphicx}
\usepackage{booktabs} 

\usepackage{url}

\usepackage{breakurl}
\usepackage[breaklinks]{hyperref}
\usepackage{bm}

\usepackage[accepted]{mlsys2025}

\usepackage{balance}
\usepackage{soul}

\def\Snospace~{\S}

\usepackage{xspace}
\newcommand{\ie}{\emph{i.e.,}\xspace}

\newcommand{\name}{PANDA\xspace}

\usepackage[font=small]{caption}
\usepackage{subcaption}
\usepackage{listings}
\usepackage{enumitem}

\usepackage{multirow} 
\usepackage{array,makecell} 

\newif\ifshowcomment
\showcommenttrue
\ifshowcomment
	\renewcommand{\comment}[1]{{\color{blue} {#1}}}
	\newcommand{\sxz}[1]{{\color{red} {#1}}}
    \newcommand{\zongpu}[1]{\textsf{\color{red}{#1}}}
    \newcommand{\qiang}[1]{{\color{blue} {#1}}}
    \newcommand{\pcomment}[1]{{\em \color{red} {#1}}}
\else
	\renewcommand{\comment}[1]{}
    \newcommand{\zongpu}[1]{}
    \newcommand{\qiang}[1]{}
    \newcommand{\pcomment}[1]{}
\fi

\newcommand{\cut}[1]{{}}

\usepackage{tikz}
\usepackage{amsmath}
\usepackage{amssymb}    
\usepackage[T1]{fontenc}
\usepackage[utf8]{inputenc}

\begin{document}

\twocolumn[
\mlsystitle{\name: Noise-Resilient Antagonist Identification in Production Datacenters}



\mlsyssetsymbol{equal}{*}

\begin{mlsysauthorlist}
  \mlsysauthor{Sixiang Zhou}{pu}
 \mlsysauthor{Nan Deng}{goo}
 \mlsysauthor{Krzysiek Rządca}{goo}
 \mlsysauthor{Xiaojun Lin}{pu}
 \mlsysauthor{Y. Charlie Hu}{pu}
\end{mlsysauthorlist}

\mlsysaffiliation{pu}{Purdue University}
\mlsysaffiliation{goo}{Google}


\mlsyskeywords{Machine Learning, Antagonist, Production Datacenter}

\begin{center}
  {\normalsize
  $^1$Purdue University
    $^2$Google Inc.
    }
\end{center}

\vskip 0.3in

\begin{abstract}

  Modern warehouse-scale datacenters commonly collocate multiple jobs
  on shared machines to improve resource utilization. However, such
  collocation often leads to performance interference caused by
  antagonistic jobs that overconsume shared resources. Existing
  antagonist-detection approaches either rely on offline profiling,
  which is costly and unscalable, or use a sample-from-production
  {approach}, which suffers from noisy measurements and
  fails under multi-victim scenarios.  We present \name, a
  noise-resilient antagonist identification framework for
  production-scale datacenters. Like prior correlation-based methods,
  \name\ uses cycles per instruction (CPI) as its performance metric,
  but it differs by (i) leveraging global historical knowledge across all
  machines to suppress sampling noise and (ii) introducing a
  machine-level CPI metric that captures shared-resource contention
  among multiple co-located tasks. 
  Evaluation on a recent Google production trace shows that \name
  ranks true antagonists far more accurately than prior
  methods---improving average suspicion percentile from 50–55\% to
  82.6\%---and achieves consistent antagonist identification under
  multi-victim scenarios, all with negligible runtime overhead.
\end{abstract}

]



\section{Introduction}
\label{Sec_intro}

Modern warehouse-scale datacenters improve resource utilization by
collocating multiple jobs from different users on the same
machine~\cite{Google_Borg,Microsoft_scheduler}. However, such
collocation often incurs performance interference, as some jobs
aggressively contend for shared resources and degrade others’
performance. This interference is especially detrimental to
latency-sensitive (LS) jobs---such as user-facing services---that must
satisfy strict quality-of-service (QoS) guarantees. Therefore, to
maximize the benefits of collocation without compromising performance,
it is crucial to identify antagonistic jobs that over-consume shared
resources and harm their co-runners.


A widely adopted approach in prior work for identifying antagonists is
{\em offline screening}~\cite{Paragon,Quasar,Smite,Bubble_flux,bubble_up}. This
method builds dedicated test environments to measure detailed
interference metrics by co-running each job with others on shared
machines. Every job is profiled in this controlled setting before
being admitted into the datacenter job pool. With such fine-grained
profiling, offline approaches can accurately identify antagonistic
behaviors and proactively mitigate interference through
scheduling. However, such an approach suffers limited scalability: modern datacenters
may host millions of jobs, making exhaustive profiling
impractical. Furthermore, job behaviors often change dynamically over
time, necessitating repeated re-profiling and exacerbating the
scalability challenge.

Another common strategy is to sample performance data directly from
production machines and identify antagonists through correlation
analysis~\cite{CPI_2,Proctor,Alibaba_Cache_antagnist}. We refer to
this as the {\em sample-from-production} approach. Most existing studies
rely on hardware performance counters, particularly cycles per
instruction (CPI), as the key metric. This choice is motivated by two
reasons:
(i) CPI can be collected using existing hardware counters with minimal software modifications, and
(ii) it closely reflects the application-level behavior of latency-sensitive (LS) jobs.
Once victims (\ie LS tasks exhibiting abnormal CPI values) are detected, prior work~\cite{CPI_2,Proctor,Alibaba_Cache_antagnist} identifies their corresponding antagonists among colocated tasks, typically through CPU-usage-based correlation analysis.

However, the sample-from-production approach suffers severely from
{\em sampling noise} in the collected performance data. The sampling program is
typically software-based and runs as a co-located job alongside
production workloads. To minimize its interference with production
performance, it must use a small sampling window, which in turn
introduces significant measurement noise. For example, our twin-task
analysis on CPI samples from a production trace~\cite{GoogleTraceV3}
reveals substantial variability across identical tasks (see detailed analysis in Section \ref{sec:problem}). Similar issues
have been reported by Proctor et al.~\cite{Proctor}, who describe this
as a distorted performance metric. Such noise can cause false victim
alarms and, more critically, incorrect antagonist
identification. Although prior
works~\cite{CPI_2,Proctor,Alibaba_Cache_antagnist} address the
false-alarm issue through various design choices, none effectively
mitigate the misidentification of antagonists.

In addition, {identifying antagonist in modern datacenter also faces
  a {\em multi-victim challenge}.}
As hardware capacity continues to increase, the number of colocated
tasks per machine has grown rapidly.
For example, \cite{CPI_2,Proctor} report an average of around
ten colocated tasks per machine, whereas recent production
traces~\cite{GoogleTraceV3} show machines hosting 50–100 tasks
simultaneously. Since these tasks share the same hardware resources,
contention can create {\em global interference effects}, causing
multiple victim tasks to appear simultaneously.
However, these victims may be affected by shared contention to different degrees. 
As a result, analyzing each victim independently, as done in~\cite{CPI_2,Proctor}, 
can lead to inconsistent antagonist identification and, consequently, confusion in diagnosis. 
An analysis of a production trace~\cite{GoogleTraceV3} further
illustrates this issue: among all machines that contain at least one
victim, 43\% host multiple victims, and reapplying the method
of~\cite{CPI_2} to any two colocated victims yields different
identified antagonists in 69\% of the cases.


To tackle both the noise problem and the multi-victim problem, we
propose \name (\textbf{P}roduction \textbf{AN}tagonist \textbf{D}etection and \textbf{A}nalysis), a
noise-resilient antagonist identification framework for
production-scale datacenters.  Like prior correlation-based
approaches~\cite{CPI_2,Proctor}, \name\ uses cycles per instruction
(CPI) as its primary performance metric, but it differs fundamentally
in scope and robustness. Instead of inferring \emph{task-level} antagonists
locally---based only on short-term observations from the same 
machine that the victim resides---
\name aggregates information globally across all machines and time windows, 
and performs analysis at the \textit{job level}.
By leveraging this global knowledge, \name produces more stable and accurate inferences, 
substantially mitigating the effects of sampling noise.

Achieving such large-scale, job-level aggregation, however, introduces
new scalability challenges.  {To address the multi-victim problem and the scalability challenge},
\name introduces a \textit{machine-level CPI}
metric that captures the overall contention level of shared hardware
resources, enabling coordinated analysis when interference
simultaneously affects multiple colocated tasks.



We evaluate \name using the large-scale Google cluster trace (v3)~\cite{GoogleTraceV3}, which contains roughly 1 M tasks executed across ~80 K production machines. The trace provides detailed per-task resource-usage and performance statistics, enabling realistic assessment of interference behaviors in real-world datacenters.

Our evaluation examines two questions:
(i) when true antagonists colocate with victims, how effectively does \name rank them compared to prior correlation-based methods~\cite{CPI_2,Proctor}?
(ii) how consistent are its antagonist predictions in multi-victim scenarios?

Across both dimensions, \name delivers substantially more accurate detection. \name assigns true antagonists an average suspicion rank of 82.6 percentile, compared to only 50–55 percentile for prior approaches, demonstrating the benefit of \name’s global view. In addition, \name achieves full consistency in multi-victim settings, whereas prior methods often disagree across victims. Finally, \name incurs negligible runtime overhead, making it practical for production-scale deployment.

\section{Motivation}
\label{sec:problem}


To fully utilize hardware resources, modern production datacenters
colocate multiple tasks on shared machines, which inevitably leads to
colocation
interference~\cite{Alibaba_datacenter,Google_Borg,Microsoft_scheduler}. Such
interference can cause significant performance degradation, especially
for latency-sensitive (LS) workloads that must meet strict
quality-of-service requirements. The key to mitigating this problem is
to accurately identify antagonists---tasks that overuse shared resources
and cause other tasks to suffer. Precisely identifying antagonists
allows datacenter operators to swiftly control interference sources
and maintain service quality.

\subsection{Limitations of Prior Approaches}

Accurately identifying antagonists in production datacenters remains challenging. Existing approaches can be broadly classified into two categories: the offline-screening approach and the sample-from-production approach.

The {\em offline-screening}
approach~\cite{Smite,Bubble_flux,bubble_up,Paragon,Quasar} builds
dedicated testing environments where tasks are executed and profiled
individually to evaluate their behavior on shared resources. Through
comprehensive task screening and profiling, these methods enable
schedulers to make informed placement decisions and proactively avoid
colocation interference. However, modern datacenters host millions of
jobs~\cite{GoogleTraceV3}, making such exhaustive profiling
prohibitively expensive. Moreover, because task behavior can change
dynamically over time, offline approaches require periodic
re-profiling, which further exacerbates their scalability limitations.

The {\em sample-from-production}
approach~\cite{CPI_2,Alibaba_Cache_antagnist,Proctor} identifies
antagonists through correlation analysis on real-time performance
samples collected from production machines. When a victim task---one
suffering from colocation interference and exhibiting abnormal
behavior---is detected, the system gathers both its performance metrics
and the resource usage statistics of colocated tasks, then computes
correlation scores to infer which task is the likely antagonist.

These methods typically rely on performance counter (PC) metrics, such
as cycles per instruction (CPI), because they can be obtained using
existing hardware with minimal software changes. Prior
work~\cite{CPI_2} shows that CPI correlates well with the
application-level performance of latency-sensitive (LS) workloads,
making it an attractive choice for interference detection.

Among online approaches, \cite{Alibaba_Cache_antagnist} collects
detailed cache metrics and constructs a contention graph across
multiple cache layers. While this design improves visibility into
cache interference, it depends on specialized cache-monitoring
hardware, making deployment across an entire datacenter
costly. Moreover, it focuses exclusively on cache contention, ignoring
other potential bottlenecks such as memory bandwidth, I/O, and
network.

In contrast, \cite{CPI_2,Proctor} infer antagonists directly from
correlation between the victim’s performance degradation and the
colocated tasks’ resource usage, assuming that antagonists’ usage
patterns fluctuate in tandem with victims’ performance. {However,
relying solely on CPI and local correlation introduces significant
vulnerability to noise: transient fluctuations in performance counters
can easily distort correlation results and lead to false or
inconsistent antagonist identification.
Further, partial observability and multi-task contention also blur
these correlations. Below, we use real-world datacenter traces to explain why the sample-from-production approach
remains fragile and prone to both false positives and incorrect
antagonist attribution.}

\subsection{Challenges in Production Environments}

\textbf{The noise problem}.  When sampling performance counter (PC)
metrics on production machines, it is necessary to prevent the
monitoring program from interfering with production workloads. To
minimize its impact, the sampling window for performance metrics is
typically very short. For example, in a production trace~\cite{GoogleTraceV3}, the reported CPI value represents the
average over a 10-second window within every 5-minute interval. As a
result, the collected samples contain substantial measurement noise,
which can trigger false victim alarms and lead to incorrect antagonist
identification. Similar issues were also reported by Proctor et
al.~\cite{Proctor}, who observed corrupted PC samples in their
dataset. To further quantify the extent of this noise, we perform a
twin-task analysis on the production trace \cite{GoogleTraceV3}.


\textbf{Twin-task analysis}. 
We observe that several machines in~\cite{GoogleTraceV3} host two or
more instances of the same job running concurrently.  These task
instances share identical binaries, CPU and memory allocations, and
colocated neighborhoods.  We refer to any such pair as \textit{twin
  tasks}.  In the absence of noise, the performance-counter samples of
twin tasks should be nearly identical, since they execute under the
same conditions.  However, our analysis reveals otherwise.

\begin{figure}[h]
    \centering
    \includegraphics[width=0.9\linewidth]{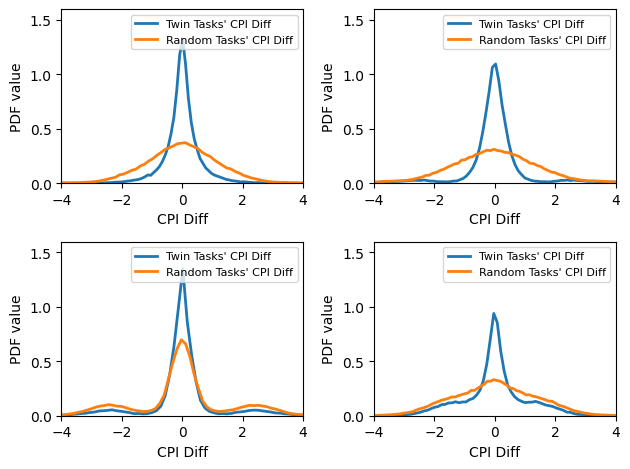}
    \caption{PDF of the absolute CPI sample difference between twin
      tasks and between random task pairs, shown for the four jobs
      with the highest number of twin-task occurrences.
    }
\label{fig_bootstrap_resampling}
\end{figure}

For each job, we compute the CPI sample difference between its twin
tasks and compare it against the CPI sample difference between two
randomly selected tasks of the same job that run on different machines
and at different times (hereafter referred to as \textit{random
  tasks}).  Although the CPI differences of twin tasks are generally
more concentrated than those of random tasks, they still exhibit
substantial variation.  Across all jobs with twin-task occurrences,
the CPI difference of twin tasks has, on average, still $0.59\times$
the standard deviation observed for random tasks of the same job.
Figure~\ref{fig_bootstrap_resampling} shows the probability density
functions (PDFs) of CPI differences for twin tasks and random tasks
among the four jobs with the most twin-task occurrences, corroborating
this observation.

In summary, while switching from random tasks to twin tasks reduces
part of the variation---reflecting the true interference signal---the
remaining variation due to noise remains dominant.  This indicates
that the signal-to-noise ratio in CPI-based inference is low.


The noise problem gives rise to the {\em incorrect antagonist identification} issue.
Neither \cite{CPI_2}
nor \cite{Proctor} addresses the impact of noise on antagonist
identification, and as a result, the inferred antagonists may be
incorrect.
Misidentifying antagonists can have serious operational
consequences. The standard recovery procedure~\cite{CPI_2,Proctor}
involves iteratively identifying the suspected antagonist, evicting
it, and then observing whether the victim’s performance improves. If
not, the process repeats with a new suspected antagonist. When noise
leads to incorrect identification, this procedure may evict multiple
innocent tasks, prolong the victim’s degradation, and severely disrupt
normal datacenter operations.

\textbf{The multi-victim problem.}
Identifying antagonists for victim tasks in modern datacenters is
further complicated by the frequent presence of \textit{multiple
  victims}, which can lead to conflicting inferences.  With the rapid
growth of hardware capacity, machines now host far more concurrent
tasks than before.  For example, prior work~\cite{CPI_2,Proctor}
reports an average of around ten colocated tasks per machine, whereas
recent production traces~\cite{GoogleTraceV3} show machines hosting
50--100 tasks simultaneously.

As a result of this
increased degree of task colocation, victims tend to appear in
clusters: the presence of one victim often indicates aggressive
shared-resource contention that can degrade multiple colocated tasks.
Indeed, an analysis of~\cite{GoogleTraceV3} shows that colocated tasks
exhibit roughly $0.7\times$ the standard deviation of CPI values
observed between randomly paired tasks, suggesting a shared
interference effect.  Consequently, multiple victims commonly arise on
the same machine, making per-task analysis both computationally
expensive and unreliable.

To further illustrate this challenge, we replicate the
correlation-based antagonist identification method proposed
in~\cite{CPI_2} using the dataset \cite{GoogleTraceV3}. Specifically,
we identify latency-sensitive (LS) victim jobs and
compute the CPU-usage–CPI correlation across all colocated tasks
within the previous hour. Tasks with the highest correlation are
considered potential antagonists, while {\em victims} are defined as
jobs whose CPI exceeds the mean by more than two standard
deviations. The results are revealing:

\begin{enumerate}
\item
  Among all per-machine, per-5-minute records containing at least one
  victim, 43\% include multiple victims;
\item
  For any two colocated victims, there is a 69\% probability that they
  accuse different antagonists.
\end{enumerate}
These findings highlight a fundamental limitation of existing
correlation-based antagonist identification approaches: when multiple
victims coexist, noise and shared contention frequently lead to
inconsistent antagonist attribution, rendering such methods unreliable
in large-scale production environments.

In summary, existing approaches on antagonist identification suffer from several limitations, motivating the need for a new solution that meets the practical requirements of production-scale datacenters. Specifically, our goal is to design an approach that:
\begin{enumerate}
    \item Scales efficiently and can be deployed seamlessly in modern production environments.
    \item
      Robustly handles measurement noise and multi-victim scenarios,
          enabling accurate antagonist identification under realistic conditions.
\end{enumerate}

\section{Key Ideas}

In this section, we present the insights and key ideas that form the
foundation of our antagonist identification framework, followed by the
main challenges that our design based on these ideas must address.

\subsection{Key Insights}

Our design is inspired by both datacenter domain knowledge and empirical observations from production traces. Specifically, we make the following observations:

\begin{enumerate}
\item
  {\bf Task similarity within a job.}  A single production job is
  typically divided into multiple tasks that execute distributedly
  across different machines. These tasks usually share the same binary
  and configuration, and therefore exhibit highly similar performance
  and resource-usage behavior.

\item
  {\bf Repetitive workload patterns.}  Datacenter workloads are highly
  repetitive over time. For example, 
  dataset~\cite{GoogleTraceV3} shows that more than 90\% of jobs have tasks that
  reappear weekly within a one-month period.
\end{enumerate}

These insights, though simple, reveal an important property of
datacenter workloads: they exhibit rich temporal and spatial
redundancy---across tasks belonging to the same job and across recurring
jobs over time. Exploiting this redundancy allows for more reliable
inference even under noisy measurement conditions, forming the
foundation of our key design principles for robust antagonist
identification:

\begin{enumerate}
\item
  {\bf Job-level inference.}
  Instead of identifying antagonistic tasks, we identify {\em antagonistic jobs}, since tasks of the same job share similar behavior and resource characteristics.
\item
{\bf Global knowledge aggregation.}  Rather than relying solely on
local information---such as recent history from a single machine---we
aggregate global information across all machines and over time to
achieve more stable and accurate inference.
\end{enumerate}

Finally, to support online monitoring in production environments, we
adopt performance counter (PC) metrics (i.e., CPI), as they can be collected
efficiently on existing hardware with minimal software overhead.

\subsection{Design Challenges}
\label{subsection_challenges}

Implementing our antagonist identification method based on the above key ideas
faces several important challenges.

\textbf{Challenge 1: Strong noise in performance counter (PC) metric
  samples.}  To avoid interfering with colocated workloads, the
sampling window for PC metrics on production machines is typically
very short. For example, in the publicly available trace~\cite{GoogleTraceV3}, the CPI metric is sampled over a
10-second window for every 5-minute record. In addition, various
external factors---such as transient network conditions, background
system activities, and dynamic business workloads---can further affect
task performance. As a result, data collected from production machines
inevitably contain strong noise, and improper denoising can easily
lead to incorrect antagonist identification. Addressing this issue
represents a primary challenge throughout this work.

\textbf{Challenge 2: Scalability of global job-level CPI analysis.}
Realizing our key ideas requires a global analysis performed at the
job level, which introduces two major scalability challenges.  
First, leveraging global knowledge demands analyzing CPI samples with
resource usage across a very large number of jobs.  
Such pairwise analysis becomes computationally prohibitive
in modern production environments.  
For example, conducting this analysis on 
\cite{GoogleTraceV3} would require computing relationship between
hundreds of thousands of LS jobs and the
resource usage of millions of jobs.

Second, obtaining joint observations for a given job pair requires
that both jobs have been colocated on the same machine during at least
one time interval---a condition determined by datacenter scheduling
policy rather than statistical sufficiency.  
As a result, data availability across job pairs can be highly uneven:
even when aggregating across the full production history, many job
pairs have insufficient co-location samples, making their antagonist
inference particularly vulnerable to noise.

\section{Design}
\label{sec:design}

In this section, we present the detailed design of our antagonist identification framework, \textbf{P}roduction \textbf{AN}tagonist \textbf{D}etection and \textbf{A}nalysis (\name).

\subsection{Features of \name}

We first describe the key design features that enable \name\ to
address the challenges discussed earlier and distinguish it from prior
approaches.

\textbf{Global view}.  Unlike prior work that infers antagonists using
only short-term data from a single machine, \name\ is designed with a
global view. It aggregates historical information from all machines
across the datacenter, enabling inference over a much broader temporal
and spatial scope. This holistic perspective allows \name\ to
effectively suppress local noise and produce more accurate and stable
antagonist identification results.

\textbf{Machine-level performance metric.}  \name\ operates at the
machine level, analyzing relations between each job’s activity and
the overall machine performance, such as the machine-wide aggregated
CPI.
Although aggregating to the
machine level may smooth out certain abnormal behaviors from
individual tasks, we consider this trade-off reasonable. When only a
small subset of tasks exhibits irregular performance, such anomalies
are likely due to non-interference factors---such as workload patterns
or transient system fluctuations---rather than true resource
contention. In contrast, genuine performance interference typically
stems from severe shared-resource contention, which affects all
colocated tasks to some degree. Consequently, the machine-level
performance metric not only mitigates the multi-victim problem
identified in Challenge 2 but also helps average out measurement
noise, addressing the false alarm problem from Challenge 1 simultaneously.

We note that aggregating performance metrics at the machine level also
changes the definition of an antagonist. In prior work, antagonists
are typically defined as tasks that degrade the performance of their
colocated tasks. In contrast, in our formulation, an {\em antagonist} refers
to a task that causes degradation in the overall machine-level
performance. This broader definition aligns with our use of aggregated
metrics and reflects the global impact of resource contention within a
shared hardware environment.

\subsection{Design Overview}

We first give an overview of the design of \name, our antagonist identification framework.

\name\ identifies antagonistic jobs by analyzing machine usage records---either from offline traces or real-time monitoring data. It uses cycles per instruction (CPI) as its primary performance metric, a well-established performance counter (PC) indicator of system-level interference in prior studies \cite{CPI_2,Proctor}.

\name\ consists of two major components: an offline phase and an online phase.

\textbf{Offline phase.}
In this phase, \name\ continuously collects performance and resource usage data from all machines across the datacenter. Specifically, for each machine and each time slot, it records:
\begin{itemize}
\item the list of tasks running on the machine,

\item the average CPU usage of each task during that time slot, and

\item the sampled CPI values of important (e.g., latency-sensitive) tasks.
\end{itemize}

Using this information, PANDA maintains an {\em antagonist score} for each
job and updates it incrementally as new data arrive. 
The score is computed by aggregating the machine-level CPI experienced
by the job’s tasks, weighted by each task’s average CPU usage. 
This weighting emphasizes periods in which a job is actively consuming
resources and thus more likely to exert interference. 
By aggregating such signals across all machines and time, PANDA
captures global interference patterns rather than relying solely on
local, short-term correlations.

\textbf{Online: Antagonist identification}.  In the online phase,
\name\ passively monitors live machines and is triggered whenever a
machine exhibits abnormal behavior---such as a CPI value significantly
higher than its baseline. Upon detection, \name\ analyzes the set of
colocated jobs running on that machine and leverages the globally
maintained antagonist scores computed during the offline phase to
determine the most likely source of interference. Specifically, among
all colocated tasks, \name\ selects the batch job with the highest
antagonist score as the predicted antagonist, allowing for quick and
informed mitigation of performance degradation.

By combining a global historical perspective with machine-level performance metrics, \name\ effectively filters out sampling noise and accurately identifies antagonists even in large, noisy production environments.

\subsection{Offline Phase Design Details}

In this subsection, we describe the detailed design of the offline phase, which consists of two main components: data preprocessing and antagonist score calculation.

\subsubsection{Preprocessing}

After collecting raw monitoring data, \name\ performs a series of preprocessing steps to transform it into model-ready input vectors. The preprocessing includes normalization, aggregation, and splitting, as detailed below.

\textbf{Normalization}.
The first step is to normalize the CPI samples. A numerically high CPI value for a task can arise from two distinct causes:
(1) Intrinsic factors, where a task is inherently computation-intensive, making a high CPI value normal for that task; and
(2) Extrinsic factors, where the CPI increase results from interference or resource contention.
\name\ focuses on the latter, as these reflect performance variability that should be minimized.

Following the method in~\cite{CPI_2}, \name\ applies \textbf{standard
  normalization} to all CPI samples. Specifically, for each job $j$,
we compute its average CPI, denoted as $\overline{\text{CPI}}_j$, and
its standard deviation {$\sigma(\text{CPI}_j)$} using all CPI samples of
its tasks. {Let the CPI sample of task $k$ {from job $j$} at time $t$  be
  $\textit{CPI}_{k,t}$.} The \textit{normalized CPI (nCPI)} for a task
$k$ from job $j$ at time $t$, denoted as $\textit{nCPI}_j\left[k, t\right]$, is
then calculated as:

\begin{equation}
    \textit{nCPI}_j[k,t] = (\textit{CPI}_{k,t} - \overline{\textit{CPI}}_j) / {\sigma(\textit{CPI}_j)}.
\end{equation}


\textbf{Aggregation}.
After normalization, we aggregate the normalized CPI values
(\textit{nCPI}) into a machine-level metric, denoted as
\textit{mnCPI}.  Specifically, \textit{mnCPI} represents the average
\textit{nCPI} of all latency-sensitive (LS) jobs
running on a given machine.  Since LS jobs account
for the most critical and performance-sensitive workloads in
production datacenters, we consider \textit{mnCPI} a meaningful
indicator of the overall machine performance.

Moreover, such jobs are present frequently in real-world deployments.
For example, in the Google trace~\cite{GoogleTraceV3}, cell~\texttt{a}, each machine hosts on average 13.5 LS
tasks, with 90.1\% of machines running at least three such tasks
concurrently.

We define \textit{mnCPI} for each machine $m$ at time $t$ as follows.
Let $K_{m,t}$ denote the set of all LS and high-priority tasks
running on machine $m$ at time $t$, {each of which is denoted by $(j,k)$, i.e., the $k$-th task from job $j$}. Then,


\begin{equation}\label{eq_mnCPI}
    \textit{mnCPI}[m,t] = \frac{\sum_{{(j,k)}\in K_{m,t}} \textit{nCPI}_j[k,t]}{|K_{m,t}|}.
\end{equation}

Unlike~\cite{CPI_2,Proctor}, which focus on per-task performance
metrics, \name\ is based on machine-level
metric \textit{mnCPI}.  This design choice effectively mitigates the
\textit{multi-victim challenge} by capturing interference at the
shared hardware level rather than from individual noisy task-measurements.


\subsubsection{Calculating Antagonist Scores}

Inspired by prior work~\cite{CPI_2,Proctor}, we design the
\textit{antagonist score} in \name\ under the assumption that a job’s
interference behavior is \textbf{positively and linearly correlated}
with its CPU utilization.  
Hence, we use the correlation between \textit{mnCPI} and the colocated job's average CPU usage to produce antagonist scores.
We now provide a precise definition of the \textit{antagonist score}.

Let $K_j$ be the set
of all tasks from job $j$ and let $MT_n$ be the set of all machine-time
pairs $[m,t]$ such that task $n\in K_j$ runs in it. Recall that $u^n_{m,t}$ is the
average CPU usage of task $n$ on $[m,t]$ and $\textit{mnCPI}[m,t]$ is as defined in \eqref{eq_mnCPI}. Then, the antagonist coefficient for each job $j$ is defined as follows:
\begin{equation}\label{eq_antagonist_coeff}
    a_j = \frac{\sum_{n\in K_j} \sum_{\left[m,t\right]\in MT_n} u^n_{m,t} *\textit{mnCPI}[m,t] }{\sum_{n\in K_j} \sum_{[m,t]\in MT_n} (u^n_{m,t})^2}
\end{equation}
Note that \eqref{eq_antagonist_coeff} can be viewed as the least-square estimate of the coefficient $a_j$ in a linear model of the form $\textit{mnCPI}[m,t]=a_j u^n_{m,t}$, assuming that all other $[m,t]$ not in $MT_n$ have $u^n_{m,t}=0$.
During the online phase, when a victim appears, say, on machine $m_c$ at time $t_c$, we then multiply each collocated task $k$'s average CPU usage with the corresponding job $j$'s antagonist coefficient $a_j$, to compute this task $k$'s antagonist score, which can be viewed as task $k$'s contribution to the performance degradation.

\section{Evaluation}
\label{sec:eval}

Evaluating antagonist identification methods, in particular establishing
reliable ground truth, poses significant challenges. Below, we first summarize
how prior studies~\cite{CPI_2,Proctor,Alibaba_Cache_antagnist}
addressed this issue,
and then describe our own evaluation methodology.

\subsection{Prior Methods for Establishing Ground Truth}

Proctor et al.~\cite{Proctor} evaluated their approach using a
simulated colocation environment, where both the workloads and
antagonists were artificially generated. While this setup allows
controlled testing, it fails to capture the complexity and
heterogeneity of real production systems, limiting its
representativeness.

In contrast, \cite{CPI_2} conducted experiments directly in Google’s
production environment. Their ground truth was established through an
ablation-based study: after throttling the identified antagonists,
they observed changes in the victim’s CPI. If the CPI decreased, the
identification was deemed correct; otherwise, it was considered
incorrect. Although this approach provides strong validation, it
involves active perturbation of production systems, which is risky and
difficult to reproduce in practice.

Finally, \cite{Alibaba_Cache_antagnist} evaluated their method on a
production trace, using human expert labeling to determine ground
truth. Considering its balance between practicality and realism, we
adopt a similar trace-based evaluation methodology in our work.

\subsection{Methodology}

We next describe our evaluation methodology.

\textbf{Human-expert {labeling}.}  
Because all data in~\cite{GoogleTraceV3} are anonymized, the true
antagonists are not directly observable.  
Therefore, similar to~\cite{Alibaba_Cache_antagnist}, we rely on
\textit{human experts} to provide ground-truth labels; specifically,
we ask Google domain experts to identify jobs they know to be
antagonists.
However, human-expert labeling is inherently \emph{one-sided}.  
Experts can confidently confirm that a job is an antagonist if they
have repeatedly observed it causing interference, but it is difficult
for them to assert that a job is \emph{not} an antagonist with the
same level of certainty.  
As a result, the available ground truth is incomplete, and standard
classification metrics such as recall cannot be directly applied.
Indeed, \cite{Alibaba_Cache_antagnist} faced the same limitation and
evaluated only precision.
Following this practice, our evaluation also focuses on precision.

\textbf{Dataset.}
Table~\ref{tab:googletracev3_stats} summarizes the statistics of {cell \texttt{a} of} the dataset~\cite{GoogleTraceV3} used in our evaluation.
The trace contains 31 days of resource-usage records collected from
more than {10,000} machines in a production Google datacenter. LS and batch jobs in \cite{GoogleTraceV3} are classified according to their priority and scheduling class: LS jobs has priority $\geq 120$ and scheduling class $\geq 2$, and the rest are batch jobs.

In our experiments, the offline phase of \name\ updates antagonist
coefficients once per day.  
Thus, whenever the online phase is triggered, \name\ relies on the
coefficients computed using data up to the previous day.

Because human-expert labeling is one-sided, we only have access to a
subset of true antagonists.  
Therefore, in our evaluation we restrict analysis to records hosting
tasks belonging to jobs within this expert-identified antagonist set.

\begin{table}[t]
\centering
\caption{Summary statistics of the dataset used in our evaluation~\cite{GoogleTraceV3}. LS: latency-sensitive.}
\label{tab:googletracev3_stats}
\begin{tabular}{l l}
\toprule
\textbf{Statistic} & \textbf{Value} \\
\midrule
Trace duration & 31 days \\
Total number of jobs & $\sim$5.2 million \\
Total task execution records & $\sim$5 billion \\
Average number of colocated tasks  & 50--100 \\
\hbox{\hspace{0.2in}} per machine    & \\
Job types & LS and batch\\
Average CPU util. per machine & 45--65\% \\
Average memory utilization  & 60--70\% \\
\hbox{\hspace{0.2in}} per machine    & \\
Sampling interval for performance & 10s/5 min \\
\hbox{\hspace{0.2in}} counters (CPI)    & \\
\bottomrule
\end{tabular}
\end{table}

\textbf{Baselines.}
To evaluate the effectiveness of \name, we compare it against representative state-of-the-art antagonist identification approaches 
that cover the two major categories of prior work:

\begin{itemize}
    \item $\textbf{CPI}^2$~\cite{CPI_2}: a correlation-based method
      that detects antagonists from real-time production data by
      analyzing the correlation between victim’s CPI and colocated
      tasks' CPU usage within a recent window. 
    As a representative \textit{sample-from-production} approach,
    $\text{CPI}^2$ serves as the most directly comparable baseline to
    \name.

\item \textbf{Proctor}~\cite{Proctor}: an enhanced variant of
  $\textbf{CPI}^2$ that analyzes the correlation between a victim’s
  CPI and the CPU usage of colocated tasks using a random subsample
  within a recent time window.  A Chi-square test is applied to ensure
  that the subsample is statistically representative. 
{As a recent \textit{sample-from-production} approach,
    Proctor provides another meaningful comparison to
    \name.}
    
\item {\textbf{\name-Local}: a localized variant of \name\ that
  restricts analysis to the full history of a \emph{single} machine
  rather than leveraging global information across the cluster.
  Comparing \name\ against \name-Local isolates the benefit of \name's
  global view and highlights its improvement in
  antagonist-identification accuracy.
}

\end{itemize}

This comprehensive comparison allows us to evaluate \name’s robustness
to noise, scalability across large traces, and ability to handle
multi-victim interference in production-scale environments.
We note that \cite{Alibaba_Cache_antagnist}
 relies on dedicated cache-monitoring
hardware, making it difficult to deploy broadly in production
datacenters; therefore, it is not included as a baseline.


\textbf{Specific setups}.
We treat a task as a victim if it belongs to a high-priority or
latency-sensitive (LS) job and exhibits $\textit{nCPI} \ge 2$.
For fair comparison, we align the online triggering conditions of
\name\ and all baselines: antagonist identification begins only when
(i) the machine-level $\textit{mnCPI}$ exceeds its 99th percentile, and
(ii) at least one victim is present for three consecutive time slots.
To ensure sufficient historical coverage for estimating \name’s
antagonist coefficients, our accuracy evaluation focuses on victim
events occurring on or after day~20 of the trace.
For baseline configuration, $\text{CPI}^2$ computes correlations using
the past two hours of data from the local machine, while Proctor
selects a random one-hour subsample from the same two-hour window,
using three victim-CPI categories and a Chi-square threshold of~1.

{ \textbf{Metrics.}  Since our ground truth is one-sided,
  classical metrics, such as precision and recall, can be
  unrepresentative.
  However, we observe that
  both \name and the baselines naturally produce
\emph{scores} for all colocated tasks and then select the task with
the highest score as the antagonist.
This scoring behavior induces a natural \emph{ranking}, \ie, the scores
provide an ordered list of colocated tasks based on their suspected
likelihood of being antagonistic.

Accordingly, we measure performance using the \emph{average percentile
rank} of all expert-identified antagonists among their colocated
tasks.  
This metric evaluates how highly a method ranks the true antagonists
within each candidate set, thereby reflecting the effectiveness of the
underlying scoring mechanism used for antagonist identification.

We note that a rank of $r$ out of $n$ will be converted to a percentile $p$ by
\begin{equation}
    p = \frac{n-r}{n-1}.
\end{equation}
}

\subsection{Evaluation Results}
\label{subsec:eval_results}

In this subsection, we present the evaluation results of \name\ on cell \texttt{a} of \cite{GoogleTraceV3}. 
{We first summarize the overall average ranking percentiles of \name and baselines, 
then analyze their performance, 
and finally discuss case studies and insights derived from our evaluation.
}

\subsubsection{Average Ranking Percentiles}
We begin by average ranking percentiles of \name\ with baselines on the suspicious of the expert-identified antagonists.
Table~\ref{tab:accuracy_comparison} reports the overall average ranking percentiles of the true antagonists suspicious level on the trace. 
\name\ achieves consistently higher percentile, demonstrating its robustness against sampling noise and multi-victim interference. Consider usually $\sim$30 colocated batch jobs, \name on average rank the true antagonists in top 5, while baselines on average give a slight-above-median rank. 

\begin{table}[ht]
\centering
\caption{average ranking percentiles of antagonist identification methods.}
\label{tab:accuracy_comparison}
\begin{tabular}{cccc}
\toprule
$\textbf{CPI}^2$ & \textbf{Proctor} & \textbf{\name}& \textbf{\name-Local} \\
\midrule
 0.543 & 0.556 &0.826 &0.602  \\
\bottomrule
\end{tabular}
\end{table}

\subsubsection{Impact of Global View}

{To evaluate the effectiveness of the global aggregation strategy, we
  compare \name\ with \name-Local, a local-only variant that uses
  machine-specific historical data. As shown in
  Table~\ref{tab:accuracy_comparison}, \name achieves higher ranking
  percentiles than \name-Local, confirming that leveraging historical
  data across all machines helps suppress noise and improve inference
  stability. Further, \name-Local obtain higher ranking percentiles
  than the baselines which are limited to only recent history,
  demonstrating the effectiveness of past history in improving
  antagonist identification.
  
  }

{
\subsubsection{Handling the Multi-Victim Scenario}

We next evaluate \name\ and the baselines under multi-victim
conditions, defined as cases where two or more victims are colocated
on the same machine at the same time.  
Table~\ref{tab:multi_vic_comparison} reports the
\textit{antagonist-consistency rate}, i.e., the probability that
pairwise colocated victims identify the same antagonist.  
By design, \name\ consistently attributes interference to the same
job across colocated victims and therefore achieves a consistency rate
of~1 in all cells.  
In contrast, the baselines exhibit varying degrees of disagreement,
highlighting their susceptibility to antagonist divergence in
multi-victim scenarios.

\begin{table}[ht]
\centering
\caption{Antagonist-consistency rate regarding to the multi-victim case of antagonist identification methods.}
\label{tab:multi_vic_comparison}
\begin{tabular}{lccc}
\toprule
 $\textbf{CPI}^2$ & \textbf{Proctor} & \textbf{\name} \\
\midrule
 0.314 & 0.376 & 1 \\
\bottomrule
\end{tabular}
\end{table}
}

\subsubsection{Summary of Findings}
In summary, our evaluation demonstrates that:
\begin{itemize}
    \item \name\ achieves higher {ranking in the antagonist score than prior correlation-based methods.}
    \item The global inference design effectively suppresses sampling noise.
    \item The machine-level performance metric enables reliable analysis in multi-victim environments.
\end{itemize}

\section{Related Work}
\label{sec:related}

A large body of prior research has explored performance isolation
between latency-sensitive (LS) jobs and background jobs in production
datacenters~\cite{Alibaba_datacenter,Bubble_flux,Caladan,CLITE,Disallow_LS_collocation,Disallow_LS_collocation_2,Heracles,Paragon,PARTIES,PerfIso,Smite,Quasar,Rubik,bubble_up,CPI_2}.
These studies can be broadly categorized into three groups.

\textbf{Reactive resource partitioning.}  
The first category, exemplified by~\cite{PARTIES,Caladan,CLITE}, focuses on reactive mechanisms that respond to contention signals by partitioning specific hardware resources---such as last-level cache (LLC) partitioning via Intel Cache Allocation Technology (CAT). 
While effective in controlled environments, these approaches depend on specialized hardware support, limiting their deployment in heterogeneous production datacenters. 
Moreover, resource partitioning technologies like CAT offer only coarse-grained control and support a limited number of partitions, making them difficult to scale to modern servers that host over 100 colocated tasks.

\textbf{Offline screening and profiling.}  
The second category, represented by~\cite{Smite,Bubble_flux,bubble_up,Paragon,Quasar,Alibaba_datacenter}, employs dedicated environments to screen jobs and profile their interference behavior. 
Each job is executed in isolation or with controlled co-runners to build an interference profile, which is later incorporated into the scheduler to guide interference-aware placement decisions. 
This approach can proactively prevent colocation interference and achieve strong performance isolation under ideal conditions. 
However, applying such methods to production-scale datacenters is prohibitively expensive---profiling millions of jobs would impose substantial overhead in both computation and storage.

\textbf{Sampling from production.}  The third category of work,
exemplified by~\cite{CPI_2,Proctor,Alibaba_Cache_antagnist}, directly
samples performance metrics from production machines to identify
performance culprits, which we refer to as \textit{antagonist jobs}.
This approach is simple, scalable, and well-suited for deployment in
production datacenters.  Once antagonists are correctly identified,
colocation interference can be quickly mitigated through targeted
actions such as task throttling or eviction.  However, sampling in
production environments typically relies on short observation windows
to avoid performance overhead, which introduces significant
measurement noise.  Such noise can severely degrade the accuracy of
antagonist identification.  None of the existing works has
systematically addressed this issue, which directly motivates our
work.

\section{Conclusion}
\label{sec:conclusion}

We presented \name, a noise-resilient and scalable framework
for identifying antagonist jobs in production datacenters.  Unlike
prior task-level approaches that rely on noisy, locally sampled data,
\name\ leverages a global view across all machines and time windows,
combining machine-level performance metrics with historical
correlation analysis to achieve robust inference.  Our evaluation on
a production datacenter trace demonstrates that \name\ significantly improves the
precision and stability of antagonist identification, even under noisy
and multi-victim conditions.  By enabling accurate and low-overhead
detection of interference sources, \name\ paves the way for more
adaptive and interference-aware scheduling in large-scale production
environments.

\section*{Acknowledgements}

This work is supported in part by NSF grant 2113893.


\bibliography{reference}
\bibliographystyle{ACM-Reference-Format}



\end{document}